\documentclass[a4paper]{panl}
\usepackage{cite}
\usepackage{wrapfig}
\usepackage{graphicx}
\usepackage{amssymb}
\usepackage{amsfonts}
\usepackage{amsmath}
\usepackage{longtable}
\usepackage{rotating}
\usepackage{lscape}
\usepackage{epsfig}
\usepackage{multirow}
\newcommand{\bfsig}{{\mbox{\boldmath$\sigma$}}}
\newcommand{\bftau}{{\mbox{\boldmath$\tau$}}}

\originalTeX

\begin{document}

\title{ Analisys of $0^-$  excitations in $^{16}$O from inelastic scattering of polarized protons of intermediate energy }
\maketitle
\authors{M.S.\,Onegin$^{a}$\footnote{E-mail: onegin@thd.pnpi.spb.ru}}
\setcounter{footnote}{0}
\from{$^{a}$\,NRC "Kurchatov institute"-PNPI}

\begin{abstract}

Comparison of the calculation of inelastic proton scattering from $^{16}$O with excitation of $0^-$ levels with $T=0,1$ with accessible experimental data at different energies of incident protons is presented. The role of antisymmetrization in reaction formalism and the manifestation of the pion condensation in nuclear are discussed. To obtain more solid conclusions on these points more experimental data are needed.

\end{abstract}
\vspace*{6pt}


\label{sec:intro}
\section*{Introduction}
The 0$^-$ excitations in nuclei are interesting because it can elucidate the reaction mechanism of proton inelastic interaction with nuclear. It requires the spin-flip in the interaction.  Impulse approximation predicts the following characteristic of polarized proton scattering using only direct term\cite{PhysRevC.44.1077, PhysRevC.27.902}:
\begin{equation}
P=A_y=0,  D_{NN}=-1,
\end{equation}
where $P$ - polarization of proton after scattering, $A_y$ - analyzing power, $D_{NN}$ - depolarization coefficient (see \cite{PhysRevC.26.727}).
When exchange terms included the following relations fulfill\cite{WONG1984299}:
\begin{equation}
P=-A_y,  D_{NN}=-1.
\end{equation}
Including the distortion in elastic channel changes the (1) for direct term, but equality (2) remains true.  
The main role in the excitation of 0$^-$ levels play tensor interaction\cite{Spin1984}.
Excitations of  0$^-$ levels in even-even nuclei are pure spin excitations. Only longitudinal spin density is non zero that is why only longitudinal part of nucleon-nucleon interaction takes part in the excitation. In momentum representation it has the following form:
$$V^l(q)=\tilde t^C_1(q)-2\tilde t^T(q). $$
Central and tensor parts of the interaction include exchange parts as well, so that:
$$\tilde  t^C_1(q)=\tilde V^C_D(q)+\tilde V^C_E(Q),
\tilde  t^T(q) = \tilde V^T_D(q) -\frac{1}{2} \tilde V^T_E(Q).$$
Isoscalar and isovector longitudinal part of the interaction have different $q$ dependence, isoscalar part have flat dependence from $q$ and isovector part is more structured\cite{Spin1984}. Excitations are dominated by tensor part mainly due to exchange term. On this reason we have a possibility to test this part of nucleon-nucleon interaction and reaction formalism for different energies of scattering proton.

There are at least two excitations 0$^-$ in $^{16}$O nuclear with energies 10.96 and 12.8 MeV which are characterized by isospins $T=0$  and $T=1$.
These levels only slightly excited in inelastic scattering of intermediate energy protons while 10.96 MeV energy level has a larger cross-section. 
Differential cross-section and analyzing power were measured for the level 10.96 MeV for protons having energies 65, 135, 200, 317 and 400 MeV in \cite{PhysRevC.30.746, PhysRevC.39.1222, PhysRevC.47.1615, PhysRevC.44.1077}. The measuring of inelastic cross-section for the level 12.8 MeV was done in \cite{PhysRevC.30.746} for protons having energy 65 MeV.
In \cite{WAKASA2006485}  the differential cross-section for level 12.8 MeV was measured for protons with energy 295 MeV. We can test the tensor part of the interaction using these excitations. 

Another interesting theme is the investigation of the precursor phenomenon of the pion condensation in nuclei which can manifest itself in 
this excitation. The quantum numbers of the excited state 0$^-  \, T=1$  is coincided with pion quantum numbers, and if pion condensate  exist in nuclear it can enhance the longitudinal spin transition density  in such transitions. Earlier this theme was investigated in\cite{WAKASA2006485}. 

\label{sec:WF}
\section*{Wave function of the excitations}

The transitions  1p$_{1/2}$ -> 2s$_{1/2}$ give the main impact for the 0$^-$ excitations in $^{16}$O. Also the transitions 1p$_{3/2}$ <-> 1d$_{5/2}$ make some contribution to the excitations. We performed the calculation of the wave functions using code NuShellX@MSU\cite{BROWN2014115}. The interaction Millener-Kurath\cite{MILLENER1975315} was used. 
We also tested the Warburton-Brown\cite{PhysRevC.46.923} interaction but the quality of experiment description with this interaction is worse, so we used Millener-Kurath wave function in our calculations. 
The spin transition densities calculated using one body matrix elements obtained using wave function of 0$^-$ excitations are shown on Fig.\ref{fig01}. It is followed from the figure that the main strenghts of the excitations is localized in the nuclei interior. The transition densities for excitations $T=0$ and $T=1$ have similar form and amplitude. For this reason we anticipate the large density dependence of the effective nucleon-nucleon 
interaction for the calculated reactor cross-section and analyzing power.

\begin{figure}[h!]
\begin{center}
\includegraphics[width=100mm]{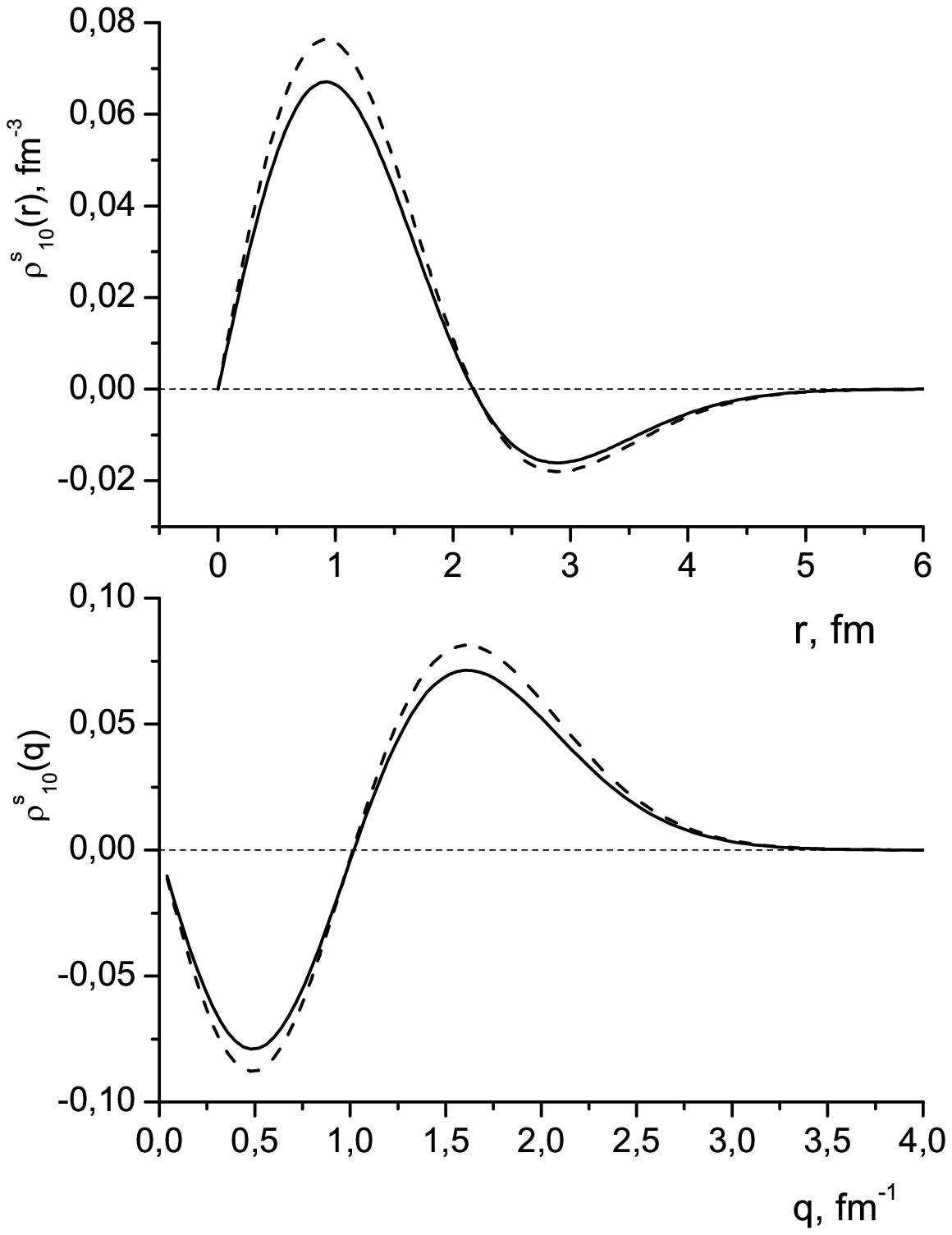}
\vspace{-10mm}
\caption{Spin transition densities for the levels 0$^-$ :  full line - $T = 0$, dash line -  $T = 1$} 
\end{center}
\labelf{fig01}
\vspace{10mm}
\end{figure}

\newpage

\label{sec:Interaction}
\section*{Effective nucleon-nucleon interaction}

The general form of the nucleon-nucleon potential used in calculations have the following form:
\begin{equation} 
V(r_{01})\ =\ V^C(r_{01}) +V^{LS}(r_{01})\,{\bf LS} + V^T(r_{01})\,S_{01}\,.
\end{equation}
The central part of the interaction in  spin and izospin space have the following components:
\begin{eqnarray}
& V^C(r_{01}) =  \\
=& V^C_0(r_{01})+V^C_\sigma(r_{01}) (\bfsig_0 \bfsig_1) + V^C_\tau(r_{01}) (\bftau_0\bftau_1)
+ V^C_{\sigma\tau}(r_{01}) (\bfsig_0\bfsig_1)(\bftau_0\bftau_1).  \nonumber
\end{eqnarray}
The $V^C_\sigma(r_{01})$ influence the izoscalar excitations channel, while $V^C_{\sigma\tau}(r_{01})$ important in izovector channel. Spin-orbit and tensor parts also have izoscalar and izovector parts. Comparison of the Fourier transformed of this components for Paris-Gamburg density dependent interaction\cite{10.1063/1.33973} is presented on Fig.\ref{fig01a} for proton energy $E_p=175$ MeV. 

\begin{figure}[h!]
\begin{center}
\includegraphics[width=100mm]{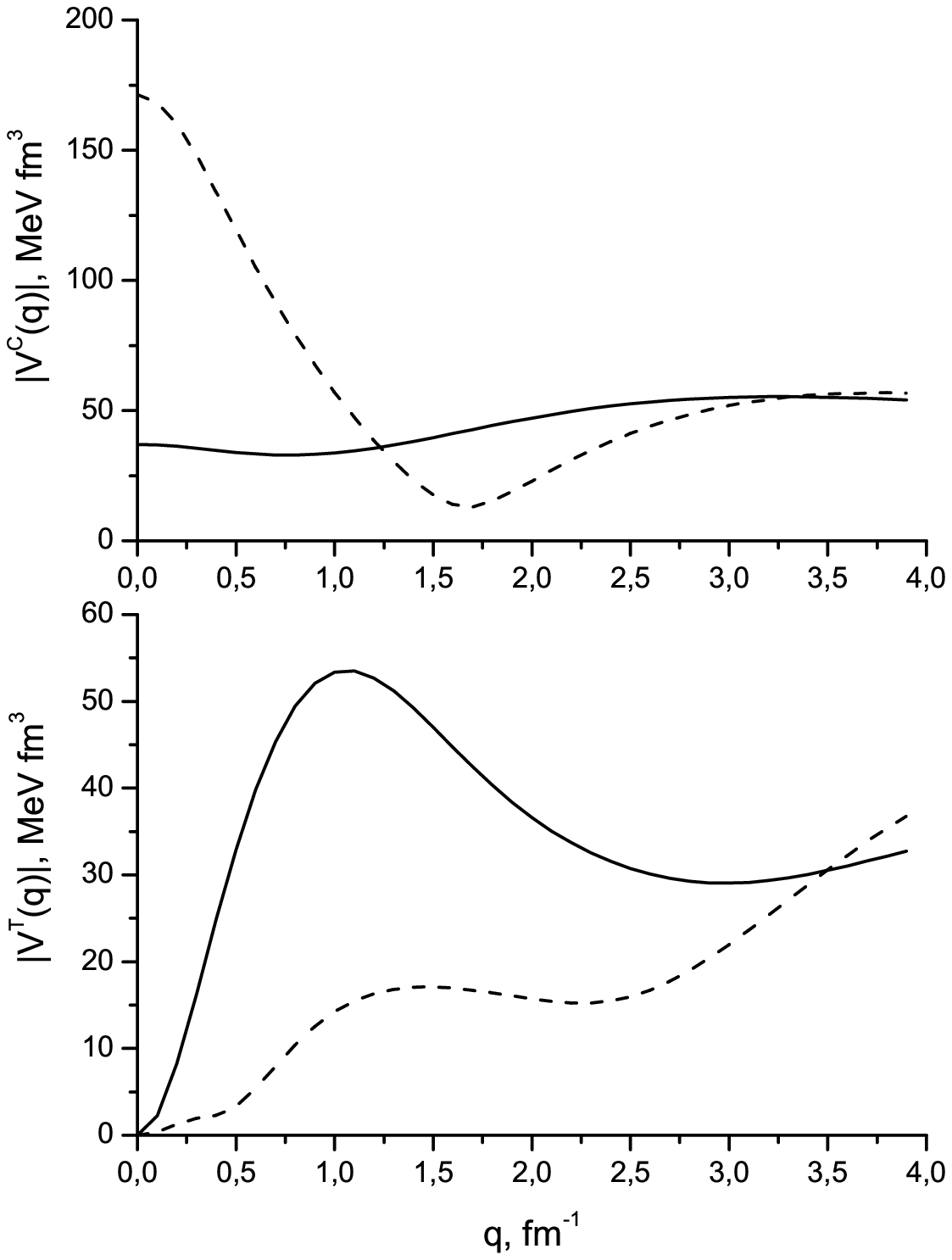}
\vspace{-10mm}
\caption{Central (upper figure) and tensor (lower figure) direct parts of the nucleon-nucleon interaction  :  full line - $T = 0$, dash line -  $T = 1$. PHDD effective interaction.}
\end{center}
\labelf{fig01a}
\vspace{10mm}
\end{figure}

\label{sec:program}
\section*{Reaction formalism}
We use DWIA formalism to calculate inelastic scattering. The Geramb and Reynal program DWBA-91\cite{DWBA91} is used. This program calculates the transition amplitude of the reaction doing full antisimmetrisation of the wave function of incident proton and nucleons in the nuclei without simplifications. Another program we have tested is LEA\cite{PhysRevC.39.1222} which also uses DWIA method but calculates the antisymmetrization in simplified manner which will be discussed shortly.

The amplitude of inelastic scattering in DWIA formalism with full antisymmetrization of wave function have a form\cite{AMOS2005230}:

\begin{align}
& T^{M_fM_i\nu'\nu}_{J_fJ_i} = \label{3} \\ 
&= \langle\chi^{(-)}_{ \nu'}(\bold{k_f}0)|\langle \Psi_{J_f M_f}(1...A)| Ag_{eff}(01)\mathcal{A}_{01}\{|\chi^{(+)}_{ \nu}(\bold{k_i}0)|\Psi_{J_i M_i}(1...A)\rangle\}, \nonumber
\end{align} 
where $\mathcal{A}_{01}$ is an antisimmetrization operator. The result of this operator is full antisimmetrizied wave function of incident proton and nucleons in target nuclear. This formula is used in DWBA-91 code. The another approach is used in LEA code where local exchange approximation is used to obtain the effective interaction of the incident proton and the struck nucleon in nuclear in local form\cite{PhysRevLett.22.895, GLENDENNING1983190}:
\begin{align}
& T^{M_fM_i\nu'\nu}_{J_fJ_i} = \label{4} \\ 
&= A \langle\chi^{(-)}_{ \nu'}(\bold{k_f}0)|\langle \Psi_{J_f M_f}(1...A)| (t_{D}(01)+t_{E}(01))|\chi^{(+)}_{ \nu}(\bold{k_i}0)|\Psi_{J_i M_i}(1...A)\rangle. \nonumber
\end{align} 
Both direct ($t_{D}$) and exchange ($t_{E}$) parts of the interaction are local density and energy dependent - $t(\rho,E,r)$, where $\rho$ is the distance between incident proton $0$ and struck nucleon $1$ in nuclear. Usually the density $\rho$ is determined as a nuclear density at mean length distance between two interact particles. 

For most of calculation we used zero-range exchange approximation\cite{PhysRevC.39.1222} option in LEA code for witch the Fourier transformed effective interaction have the following form:
\begin{equation}
t_{D}(01)+t_{E}(01))=\eta t_{NN}(q,k_A),
\end{equation}
where $\eta$ is a multiplier with value near unity\cite{PhysRevC.24.1073} and $k_A$ is the projectile wave number in the NA center of mass.
The t-matrix of the nucleon-nucleon interaction is obtained from the potential of nucleon-nucleon interaction using  formula\cite{PhysRevC.24.1073}:
\begin{equation}
t_{NN}(q,Q)=\int d\bold{r_{01}} e^{-i \bold{k_f} \bold{r_{01}}} V_{12}(\bold{r_{01}}) (1-X)  e^{i \bold{k_i} \bold{r_{01}}},
\end{equation}
where $X$ - is the exchange operator, $\bold{q}=\bold{k_i}-\bold{k_f}$ and  $\bold{Q}=\bold{k_i}+\bold{k_f}$.

We used the Paris-Gamburg density dependent (PGDD) interaction\cite{10.1063/1.33973} as effective interaction of protons with nucleons in nuclei ($V_{12} \equiv g_{eff}(01)$ from Eq. (5)). It approximates $G-$matrix of nucleon-nucleon scattering in nuclear matter. Also for comparison at incident proton energy 200 MeV we use Nakayama and Love (NL)\cite{PhysRevC.38.51} density dependent effective interaction. The comparison of this interaction with PGDD interaction was done in\cite{onegin2005testingvariousparametrizationseffective}. 

\label{sec:Isoscalar}
\section*{Isoscalar excitations}

We have calculated inelastic cross-section and analyzing power for level 0$^-$ $T=0$ for protons having energies  65, 135, 200 and 317 MeV. The PGDD interaction and DWBA-91 code was used. The elastic scattering and distorted waves were calculated using folding model and PGDD interaction. The ground state density was calculated using oscillator model wave functions with oscillator lengths 1.73 fm and full occupation of the oscillator shells up to Fermi level. The comparison of the result of the calculations with experimental data is presented on Figs.\ref {fig02},\ref{fig03},\ref{fig04} and \ref{fig05}.


\begin{figure}[hp!]
\begin{center}
\includegraphics[width=150mm]{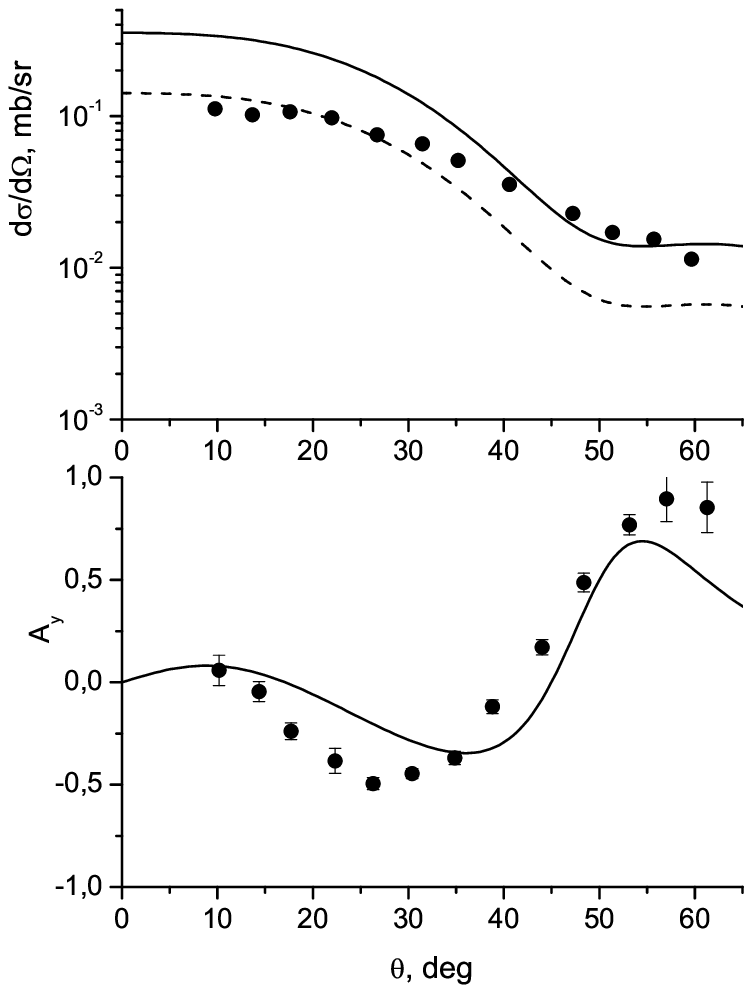}
\vspace{-60mm}
\caption{ Differential cross-section (upper part) and analyzing power
 (lower part) \\
 for level 0$^-$. Initial proton energy - 65 MeV. 
 Dash line in cross-section figure is the calculated cross-section multiplied by the factor 0.4.} 
\end{center}
\labelf{fig02}
\vspace{25mm}
\end{figure}

\begin{figure}[hp!]
\begin{center}
\includegraphics[width=150mm]{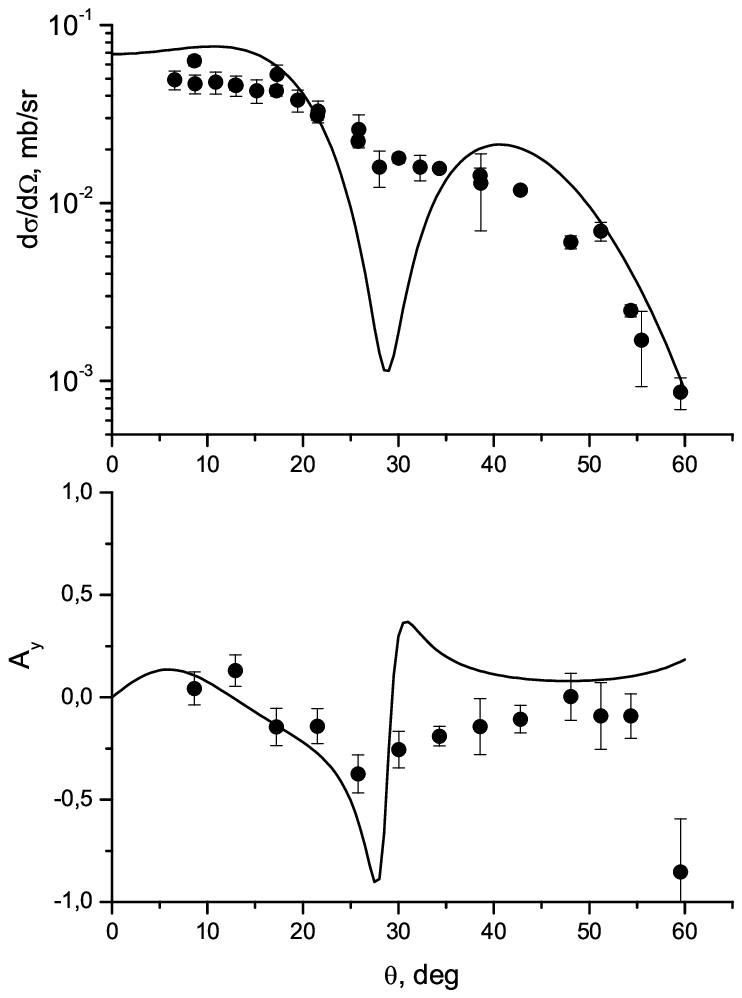}
\vspace{-70mm}
\caption{Calculated differential cross-section (upper part) and analyzing power (lower part) for level 0$^-$, $T=0$ in comparison with experimental data. Initial proton energy - 135 MeV.}
\end{center}
\labelf{fig03}
\vspace{5mm}
\end{figure}

\begin{figure}[hp!]
\begin{center}
\includegraphics[width=100mm]{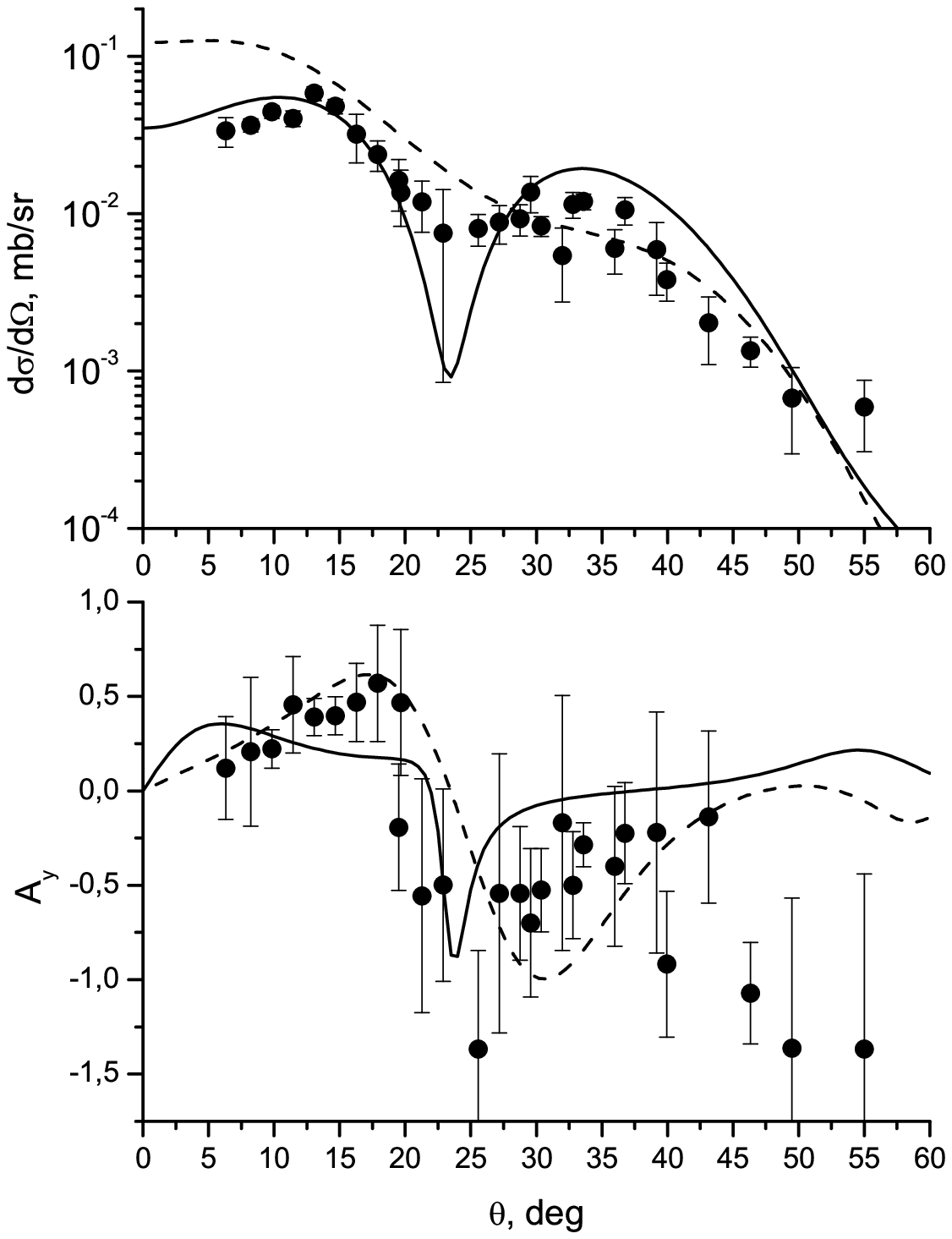}
\vspace{-10mm}
\caption{Calculated differential cross-section (upper part) and analyzing power
 (lower part) 
 for level 0$^-$ in comparison with experimental data. Initial proton energy - 200 MeV. Solid line - PGDD interaction,
 dash line - NL interaction}
\end{center}
\labelf{fig04}
\vspace{5mm}
\end{figure}

\begin{figure}[hp!]
\begin{center}
\includegraphics[width=150mm]{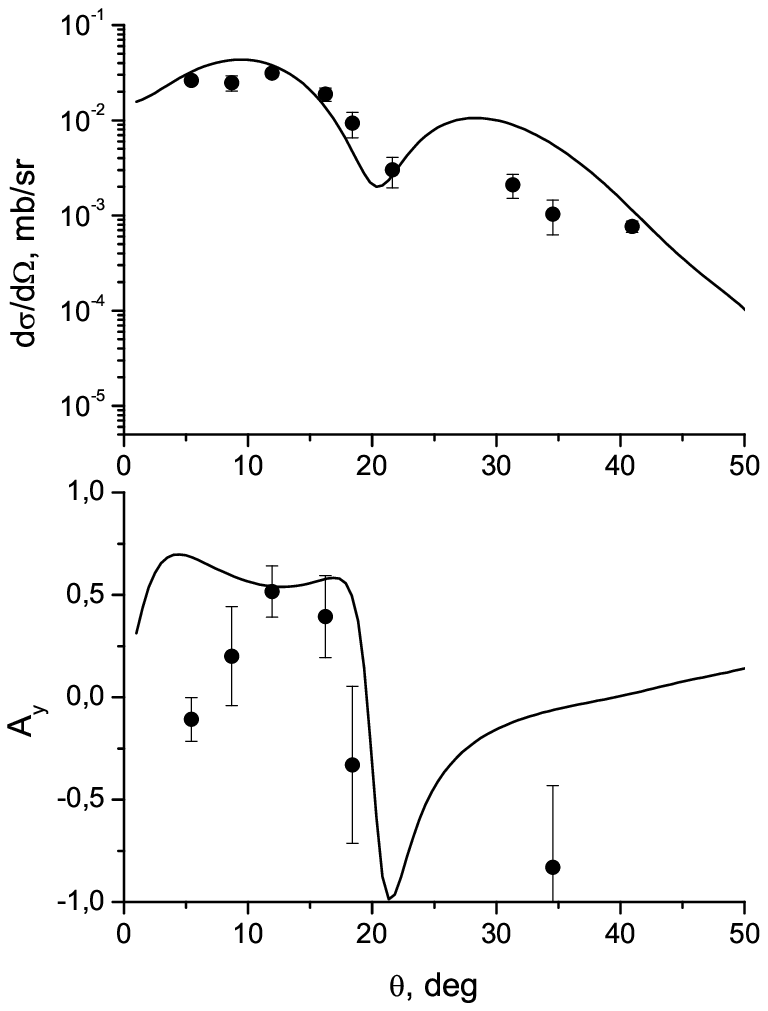}
\vspace{-70mm}
\caption{Calculated differential cross-section (upper part) and analyzing power (lower part) for level 0$^-$, $T=0$ in comparison with experimental data. Initial proton energy - 317 MeV.}
\end{center}
\labelf{fig05}
\vspace{5mm}
\end{figure}

Calculated cross-section for proton energy 65 MeV is larger than experimental data for forward angles about 2.5 times. Also the analyzing power agrees with experimental data only in a quality manner. The DWIA formalism is not sound for this energy of protons so we can state only that agreement with experiment is not bad.

Incident energy of protons 135 MeV is more valid to apply DWIA approximation. Calculated analyzing power agrees more or less with experimental data for scattering angles less than 30 degrees. The calculated cross-section exceeds experimental data for forward angles in 1.4 times but experimental data don't have a big dip in cross-section at 30 degrees. 

Comparison of the experiment and calculations is more informative for incident proton energy 200 MeV. The calculated cross-section describes experimental data up to 30 degrees scattering angle rather good here and experimental data at 22.5 degrees doesn't disagree with calculations radically. Agreement of calculated analyzing power with experiment is not satisfactory, and experimental data here is not so good as well. Alternatively for this energy we used the NL interaction in DWIA calculations using DWBA-91 code. Agreement of calculations and experiment is better for analyzing power when we use this interaction. The cross-section is described using NL interaction without maximum at 15 degrees. At the same time calculations better describe experimental cross-section and analyzing power at angles in the range 25 -50 degrees. 

For incident proton energy 317 MeV the disagreement between calculated analyzing power and experiment at forward angles enlarged but the calculated cross-section nearly describes experiment up to 20 degrees. 

\newpage

\label{sec:Isovector}
\section*{Isovector excitations}
We have calculated inelastic cross-section and analyzing power for level 0$^-$ $T=1$ for protons with incident energies  65 and 295 MeV. The PGDD interaction and DWBA-91 code was used. The comparison of the calculation with experiment is shown on Figs.\ref {fig06},\ref{fig07}.

On Fig.\ref {fig06} calculated cross-section is comparing with experimental data for incident proton energy 295 MeV. Unfortunately the experimental data exist only for large scattering angles (momentum transfer). To fit the experimental data the normalization coefficient 0.4 of the calculated cross-section is needed. Normalized cross-section describes experimental data well. The reason of using normalization  is not clear. Probably the used spin transition density for this excitation is too large. 

The comparison of the calculated cross-section with experimental data is made on Fig.\ref{fig07} for proton energy 65 MeV. Calculated cross-section surpass the experiment for forward angles and angles in the range 45 - 60 degrees. If we use the same normalization coefficient 0.4 as for incident proton  energy 295 MeV the agreement with experiment for this angles improved but there will be disagreement with experiment in the angle range 25 - 35 degrees. This angles correspond to momentum transfer to nuclear during transition about 0.85 fm$^{-1}$. At the same time the pionic enhancement should manifest itself in the cross-section enhancement for the momentum transfer about 1.7 fm$^{-1}$. Unfortunately normalized cross-section describes near perfectly the experimental cross-section at this momentum transfer for the incident proton energy 295 MeV (see Fig.\ref{fig06}). So we doesn't see any manifestation of precursor phenomenon of the pion condensation in the available experimental data.

The calculation describe the experimental data for analyzing power for proton energy 65 MeV more or less satisfactory (see Fig.\ref{fig07}). 

\begin{figure}[hp!]
\begin{center}
\includegraphics[width=200mm]{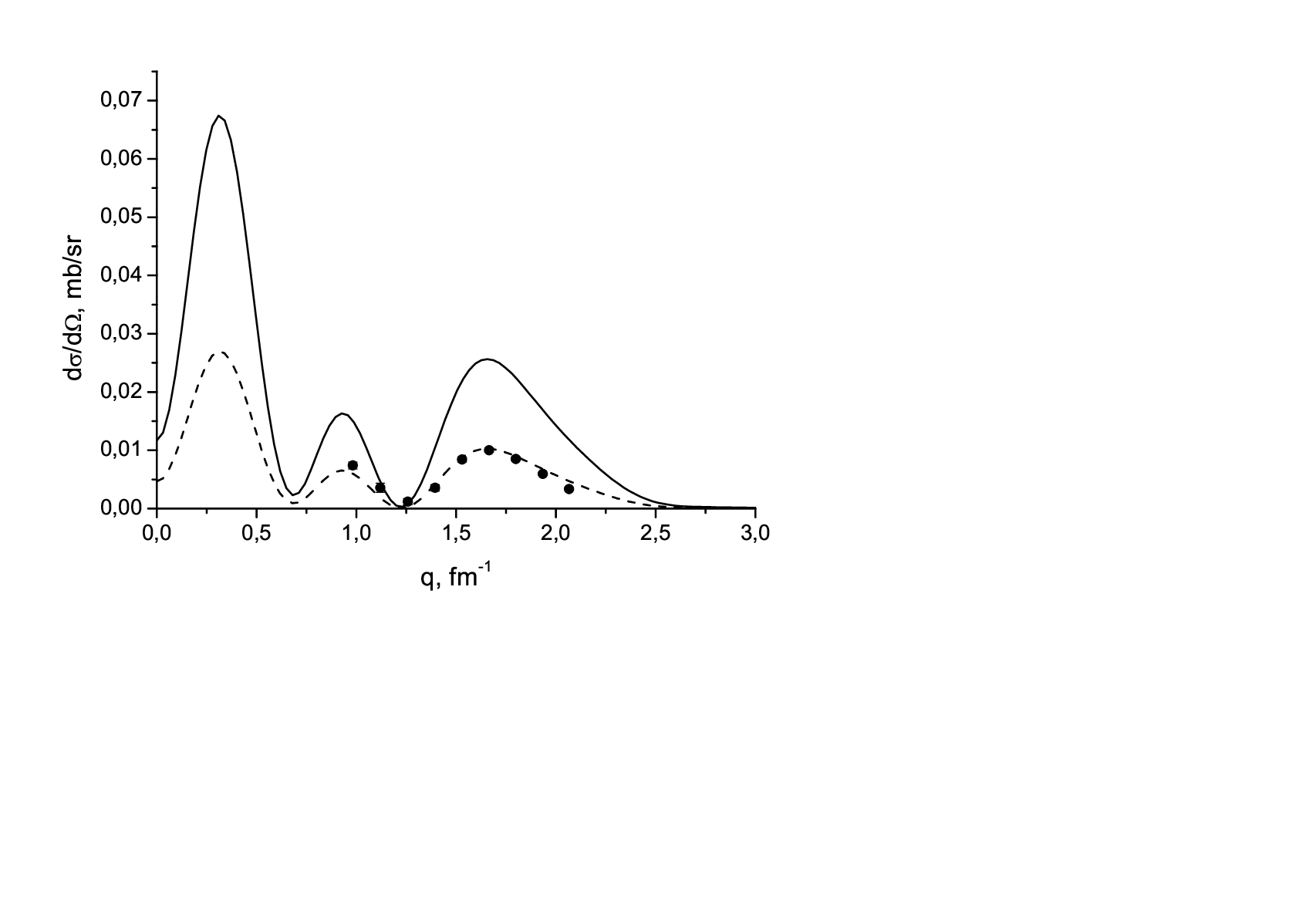}
\vspace{-40mm}
\caption{Calculated differential cross-section for level 0$^-$, $T=1$ in comparison with experimental data. Initial proton energy - 295 MeV. The dash line is the calculated cross-section multiplied by the factor 0.4.}
\end{center}
\labelf{fig06}
\vspace{5mm}
\end{figure}

\begin{figure}[hp!]
\begin{center}
\includegraphics[width=150mm]{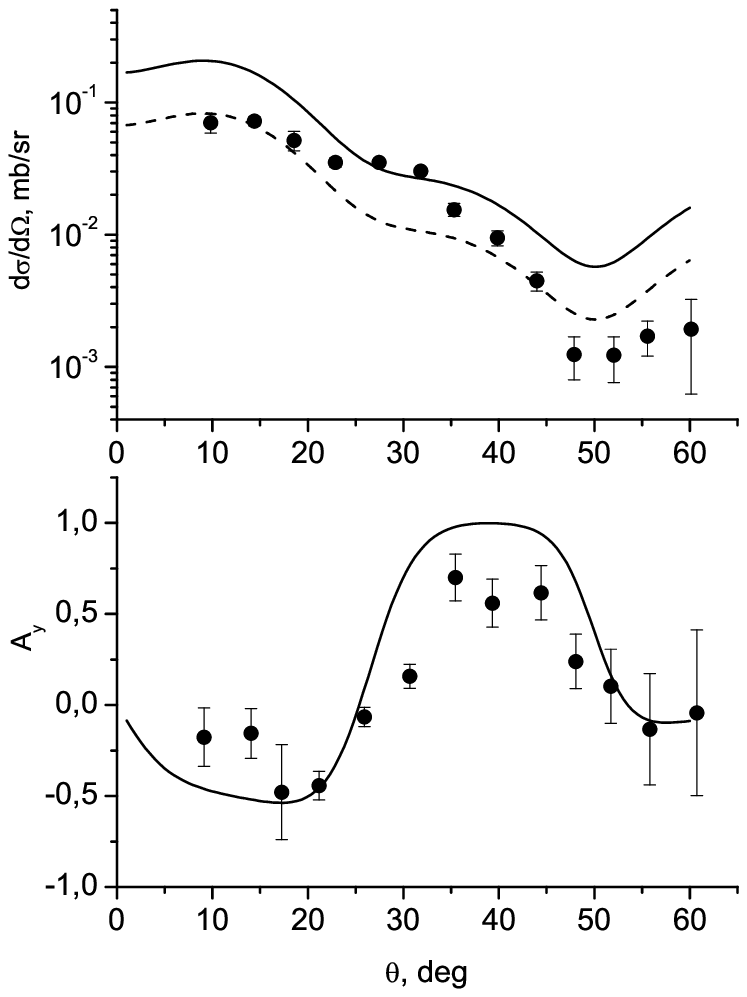}
\vspace{-80mm}
\caption{Differential cross-section (upper part) and analyzing power
 (lower part) \\
 for level 0$^-$ and $T=1$. Initial proton energy - 65 MeV. 
 Dash line in cross-section figure is the calculated cross-section multiplied by the factor 0.4.}
\end{center}
\labelf{fig07}
\vspace{5mm}
\end{figure}

\newpage

\label{sec:Antisymmetrization}
\section*{The role of antisymmetrization}

To elucidate the role of antisymmetrization on the cross-section and analyzing power we compare the DWBA-91 and LEA programs calculation with experimental data. On Figs.\ref{fig08},\ref{fig09} the comparison of DWBA-91 and LEA calculations with experimental data for 135 MeV and 200 MeV incident proton energy is done. The difference between calculated cross-section using two programs is not large and is only significant at forward angles smaller then 5 degrees where the experimental data are absent. On the other hand the difference between calculated analyzing powers is more solid. Meanwhile the experimental data for analyzing power  have larger errors and there is also lucking of experimental data at forward angles where the differences between calculation results using two programs is especially large. 
The different picture we saw for incident proton energy 400 MeV. The comparison of the two calculations with experimental data\cite{PhysRevC.44.1077} is shown on Fig.\ref{fig10}. From the figure it followers that the description of the cross-section described using two programs have nearly the same quality but the description of the analyzing power with LEA code is much better.
\begin{figure}[htp!]
\begin{center}
\includegraphics[width=100mm]{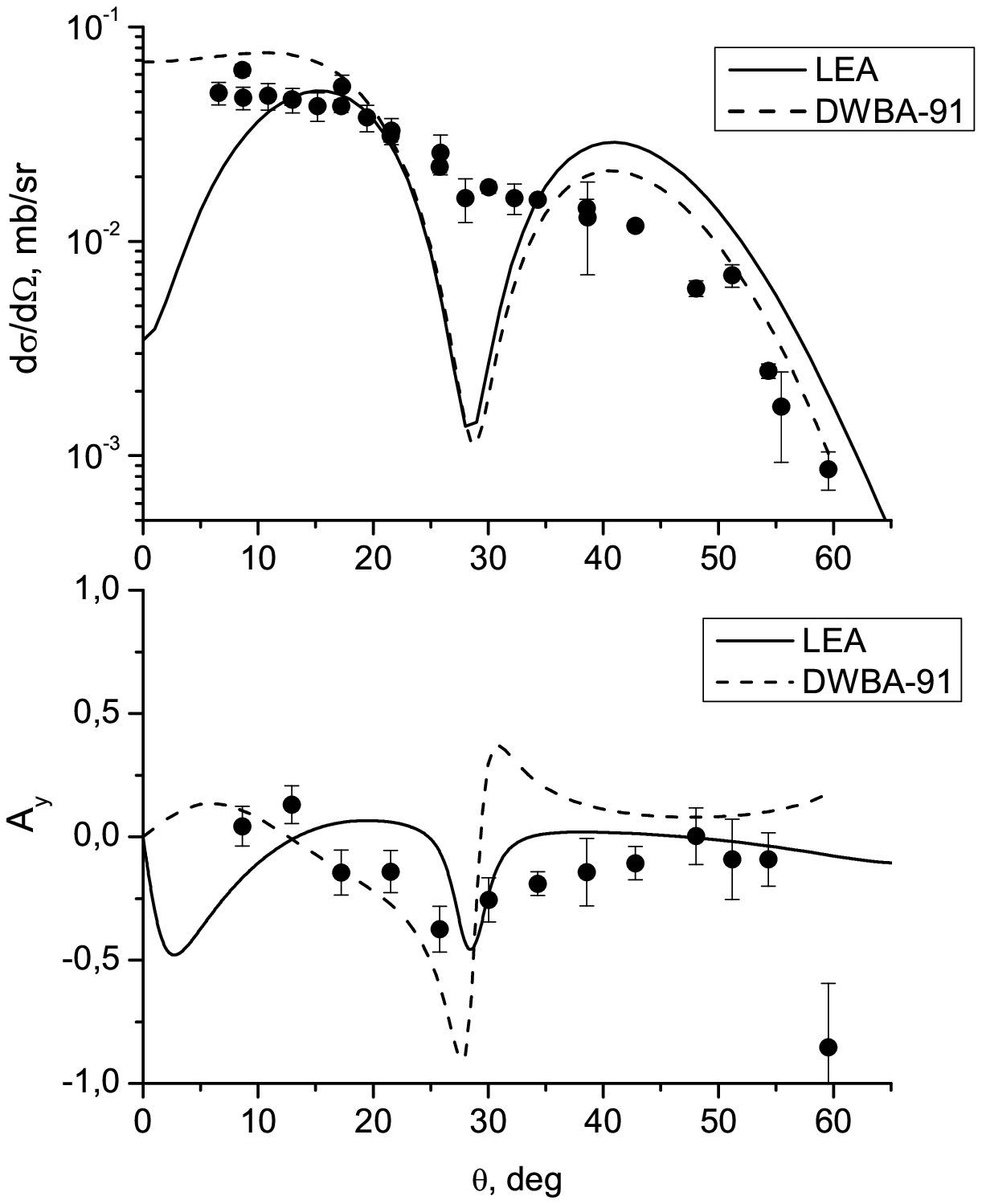}
\vspace{-10mm}
\caption{Differential cross-section (upper part) and analyzing power
 (lower part) \\
 for level 0$^-$ and $T=0$. Initial proton energy - 135 MeV. 
 Dash line - DWBA-91 calculations; solid line - LEA calculations.}
\end{center}
\labelf{fig08}
\vspace{5mm}
\end{figure}

\begin{figure}[htp!]
\begin{center}
\includegraphics[width=100mm]{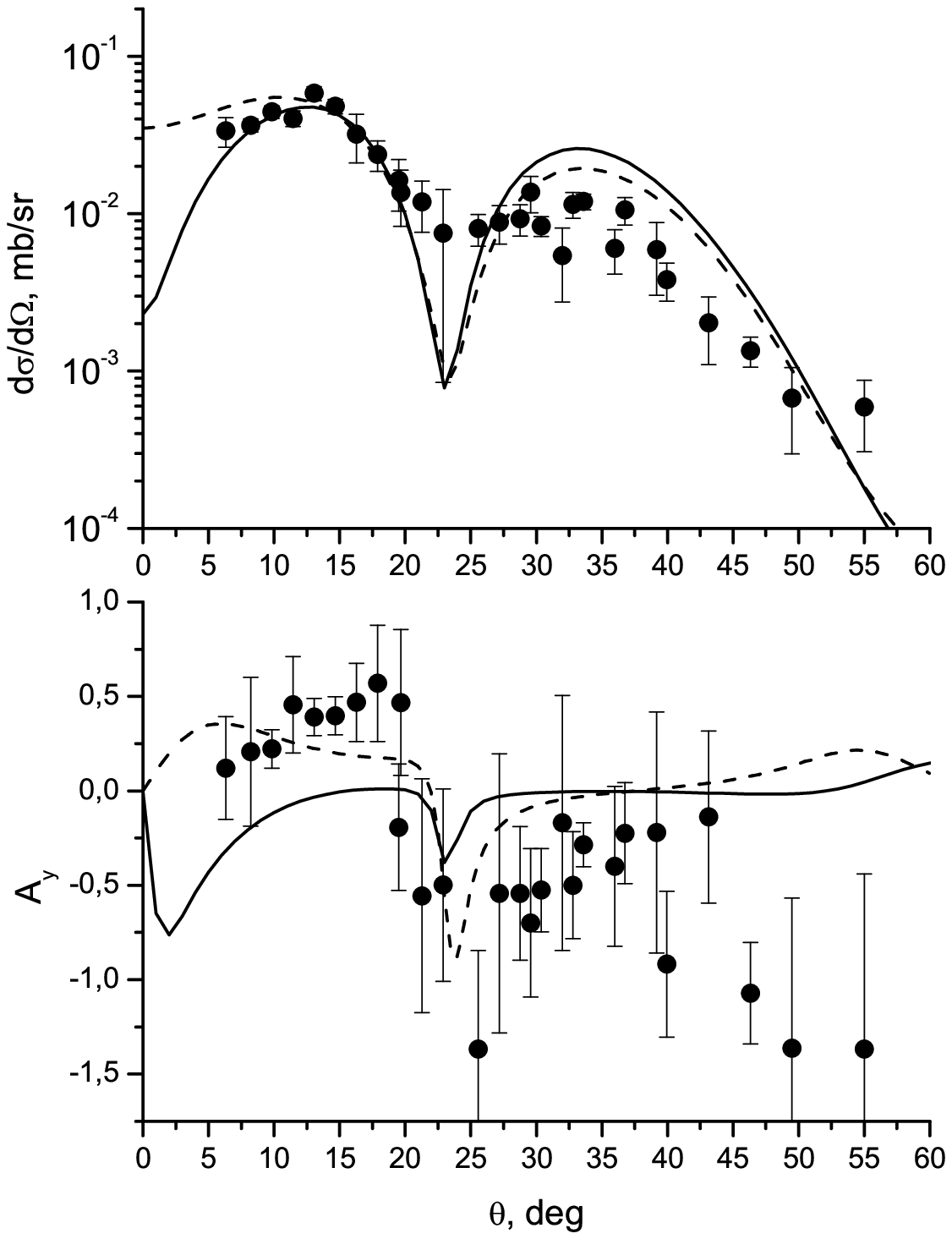}
\vspace{-10mm}
\caption{Differential cross-section (upper part) and analyzing power
 (lower part) \\
 for level 0$^-$ and $T=0$. Initial proton energy - 200 MeV. 
 Dash line - DWBA-91 calculations; solid line - LEA calculations.}
\end{center}
\labelf{fig09}
\vspace{5mm}
\end{figure}

\begin{figure}[htp!]
\begin{center}
\includegraphics[width=100mm]{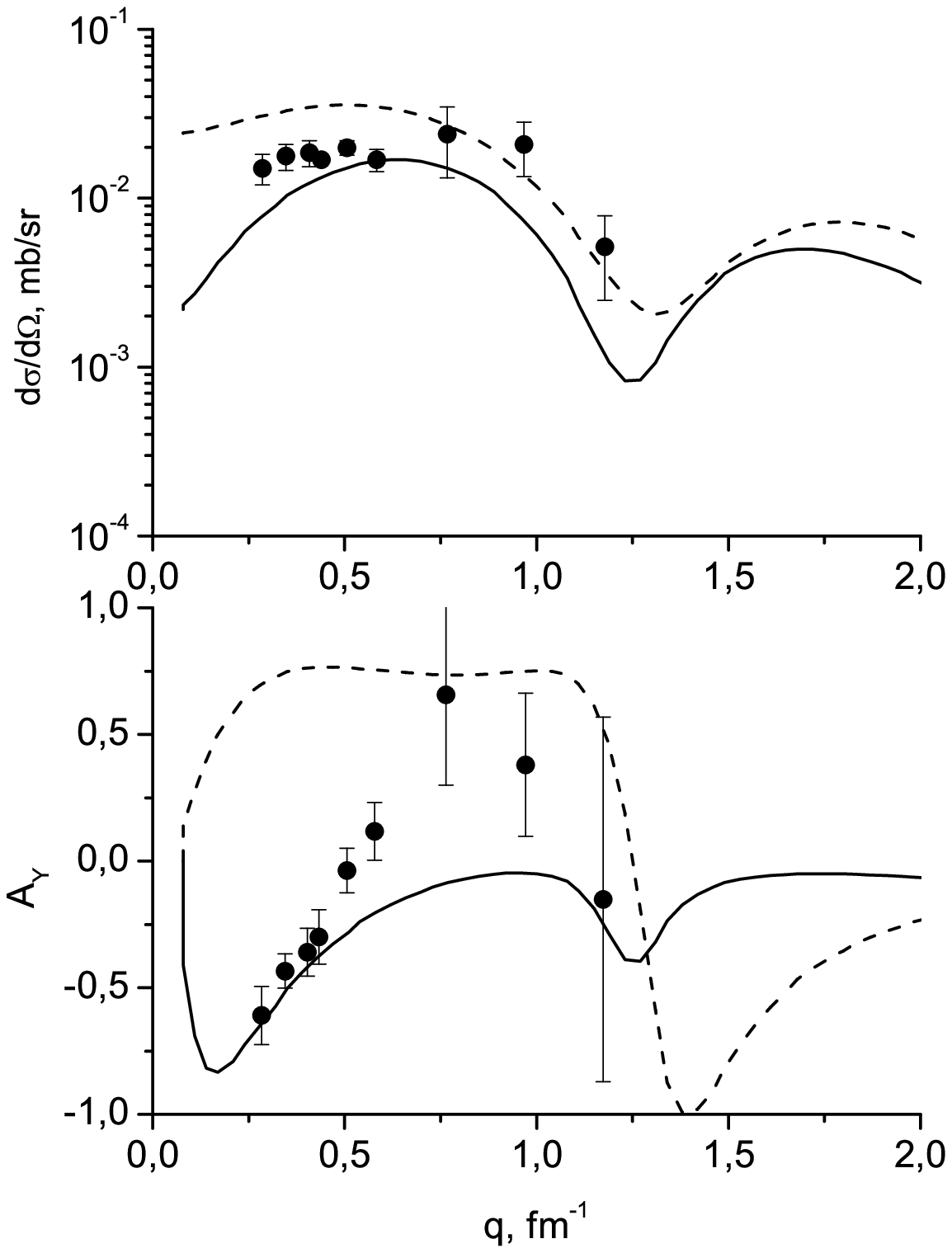}
\vspace{-10mm}
\caption{Differential cross-section (upper part) and analyzing power
 (lower part) \\
 for level 0$^-$ and $T=0$. Initial proton energy - 400 MeV. 
 Dash line - DWBA-91 calculations; solid line - LEA calculations.}
\end{center}
\labelf{fig10}
\vspace{5mm}
\end{figure}

\section*{Summary and conclusions}
Calculations of the cross-section and analyzing power for $0^-$ energy levels by protons from $^{16}$O for different energies of incident protons were done and compared with experimental data. The DWBA-91 and LEA code which uses DWIA formalism of the reaction theory were used. The difference between this programs is that DWBA-91 uses full antisymmetrized  wave function of incident proton and nucleon in nuclear while LEA code uses antysymmetrization only between incident proton and struck nucleon. As it follows based on accessible experimental data there is no advantages of one or another approach. 

We didn't find the traces of the pion condensate in nuclei using cross-section calculations of $0^-, T=1$ excitation by 295 MeV protons from $^{16}$O nuclei. The additional experimental data needed for smaller momentum transfer to prove this more solid.

\newpage

\label{sec:acknowledgement}
\section*{Acknowledgement}

The author grateful to  A.V. Plavko for discussions.


\bibliographystyle{pepan}
\bibliography{pepan_biblio}

\end{document}